\documentclass[pre,eqsecnum,showpacs,amssymb]{revtex4}

\usepackage{graphicx}        
\usepackage{amsmath}
\makeindex             

\begin{document}

\title{Polymer Chain Adsorption on a Solid Surface: Scaling Arguments and
Computer Simulations}
\author{A. Milchev$^{1,2}$, V. Rostiashvili$^2$, S. Bhattacharya$^2$
and T. Vilgis$^2$}
\affiliation{Institute of Physical Chemistry, Bulgarian Academy of Sciences,
1113 Sofia, Bulgaria, \texttt{milchev@ipc.bas.bg} and Max Planck Institute for
Polymer Research, 10 Ackermannweg 55128 Mainz, Germany}

\begin{abstract}
We examine the phase transition of polymer adsorption as well as the underlying
kinetics of polymer binding from dilute solutions on a structureless solid
surface. The emphasis is put on the properties of regular multiblock copolymers,
characterized by block size $M$ and total length $N$ as well as on random
copolymers with quenched composition $p$ of sticky and neutral segments. The
macromolecules are modeled as coarse-grained bead-spring chains subject to a
short-ranged surface adhesive potential. Phase diagrams, showing the variation
of the critical threshold for single chain adsorption in terms of $M$ and $p$
are derived from scaling considerations in agreement with results from computer
experiment.

Using both scaling analysis and numerical data from solving a system of coupled
Master equations, we demonstrate that the phase behavior at criticality,
and the adsorption kinetics may be adequately predicted and understood, in
agreement with the results of extensive Monte Carlo simulations. Derived
analytic expressions for the mean fraction of adsorbed segments as well as for
Probability Distribution Functions of the various structural building blocks
(i.e., trains, loops, tails) at time $t$ during the chain attachment process are
in good agreement with our numeric experiments and provide insight into the
mechanism of polymer adsorption.
\end{abstract}
\maketitle


\section[]{Introduction}
\label{sec:intro}

The adsorption of polymers on solid surfaces is a long-standing problem which
plays an important role in a number of applications in technology (protective
coatings, adhesives, lubricants, stabilization of colloids, flocculation, etc.)
and biology (adsorption of biopolymers, etc.). As a phenomenon it poses a number
of challenging scientific problems~\cite{Wu,Eisenriegler,Sanchez,Netz} too.
Important theoretical contributions have been made by Birshtein\cite{Birshtein},
de Gennes\cite{deGennes}, Eisenriegler et al.\cite{Eisen}. Later studies have
covered adsorption of polyelectrolytes~\cite{Yamakov}, dynamics of adsorbed
chains~\cite{AMKB_Macromolecules1996} and adsorption on chemically heterogeneous
surfaces~\cite{Balasz1}. The close relationship between analytic theory and
computer experiments in this
field~\cite{AM_Dekker,Balasz1,Sommer,Metzger,Grassberger} has proved especially
fruitful and instructive.

While the investigations mentioned above have been devoted exclusively to
homopolymers, the adsorption of copolymers (e.g., random or multi-blocks
copolymers) still poses open questions. Thus, for instance, the critical
adsorption potential (CAP) dependence on block size $M$ at fixed concentration
of the sticking $A$-monomers is still unknown as are the scaling properties of
{\em regular multi-block copolymers} in the vicinity of the CAP. From the
theoretical perspective, the case of diblock copolymers has been studied by
means of the Grand Canonical Ensemble (GCE)-approach~\cite{Zhulina,DiMarzio},
within the Self-Consistent Field (SCF)-approach~\cite{Evers,Fleer}, or by Monte
Carlo computer simulations~\cite{Balasz2,Zheligovskaya}.  The case of {\em
random copolymers} adsorption has gained comparatively more attention by
researcher so far. It has been investigated by Whittington et
al.~\cite{Whit_1,Whit_2} using both the annealed and quenched models of
randomness. The influence of sequence correlations on the adsorption of random
copolymers has been treated by means of the variational and replica method
approach\cite{Polotsky}. Sumithra and Baumgaertner~\cite{Baum} examined the
question of how the critical behavior of random copolymers differs from that of
homopolymers. Thus, among a number of important conclusions, the results of
Monte Carlo simulations demonstrated that the so called adsorption (or,
crossover) exponent $\phi$ (see below) is independent of the fraction of
attractive monomers $n$.

The adsorption kinetics of polymers has been intensively studied both
experimentally~\cite{Vincent,Takahashi} and
theoretically~\cite{Konst,Shaffer,O'Shaughnessy_1,O'Shaughnessy_2,Descas,%
Ponomar,Panja} since more than two decades now. A key parameter thereby is the
height of the free energy adsorption barrier that the polymer chain has to
overcome so as to bind to the surface. High barriers are usually referred to
as cases of {\em chemisorption} as opposed to those of {\em physisorption} which
are characterized by low barriers for adsorption. Depending on the strength of
the binding interaction $\epsilon$, one distinguishes then between weak
physisorption when $\epsilon$ is of the order of the thermal energy $k_BT$
(with  $k_B$ being the Boltzmann constant), and strong physisorption when
$\epsilon \ge 2k_BT$. One of the important questions concerns the scaling of the
adsorption time $\tau_{\rm ads}$ with the length of the polymer chain $N$ in
dilute solutions. For homopolymers in the regime of strong physisorption (that
is, for sticking energy considerably above the CAP) computer
experiments~\cite{Shaffer,Descas,Ponomar} suggest $\tau_{\rm ads} \propto
N^\alpha$ where $\alpha$ is related to the Flory exponent $\nu$ as $\alpha = 1 +
\nu \approx 1.59$. This result follows from the assumed {\em zipping} mechanism
in the absence of a significant barrier whereby the chain adsorbs predominantly
by means of sequential, consecutive attachment of monomers, a process that
quickly erases existing loops on the substrate. For the case of {\em weak}
adsorption, one should mention a recent study~\cite{Panja}, where
one finds in contrast $\alpha = (1+2\nu)/(1+\nu) \approx 1.37$ which suggests
shorter time scale for surface attachment. In chemisorption, the high barrier
which attaching monomers encounter slows down the binding to the surface, the
chain gains more time to attain equilibrium conformation and the adsorption
process is believed to involve large loop formation giving rise to {\em
accelerated zipping} mechanism~\cite{O'Shaughnessy_1,O'Shaughnessy_2}. The
predicted value of $\alpha$ in agreement with MC results is $\alpha \approx 0.8
\pm 0.02$~\cite{Ponomar}. A comprehensive overview of experimental work and
theoretic considerations may be found in the recent review of O'Shaughnessy and
Vavylonis~\cite{O'Shaughnessy_2}.

In the present contribution we focus on copolymer physisorption, extending thus
the aforementioned studies of homopolymers statics and kinetics. We show how
scaling analysis as well as different MC-simulation methods help understand the
critical behavior of multi-block and random copolymers. It turns out that the
critical behaviour of these two types of copolymers can be reduced to the
behavior of an effective homopolymer chain with "renormalized" segments. For
multi-block copolymers one can thus explain how the adsorption threshold depends
on the block length $M$ and even derive an adsorption phase diagram in terms of
CAP against $M$. In the case of random copolymers, the sequence of sticky and
neutral (as regards the solid substrate) monomers within a particular chain is
usually fixed which exemplifies a system with {\em quenched randomness}.
Nevertheless, close to criticality the chain is still rather mobile, so that the
sequence dependence is effectively averaged over the time of the experiment and
the problem can be reduced to the easier case of {\em annealed randomness}. We
show that the MC-findings close to criticality could be perfectly described
within the annealed randomness model.

For both regular multiblock as well as for random copolymers, we compare the
predicted kinetics of adsorption in the regime of strong physisorption, to
consistent numeric data derived from simulations and coupled Master equations.
We demonstrate that the observed adsorption kinetics is close to that of
homopolymers and suggest interpretation of typical deviations. Eventually, we
should like to stress that the complex polymer hydrodynamics near an interface
has remained beyond the scope of this paper.

\section[]{Simulation Methods}
\label{sec:MC}

Apart from the frequently used Bond-Fluctuation Method
(BFM)~\cite{Shaffer,Ponomar}, two coarse-grained models, a bead-spring
off-lattice model~\cite{AMKB_Macromolecules1996} and a cubic lattice
model implementing the so called pruned-enriched Rosenbluth method
(PERM)~\cite{Grassberger}, are used to test theoretical predictions.

\subsection[]{Off-lattice bead-spring model}
\label{off-lattice_MC}

In our computer simulations we use a coarse grained off-lattice bead spring
model\cite{AM_Dekker} to describe polymer chains. The system consists of a
single chain tethered at one end to a flat structureless surface so as to avoid
problems with translational entropy depending on the box size. There are two
kinds of monomers: "A" and "B", of which only the "A" type feels an attraction
to the surface. The surface interaction of the "A" type monomers is described by
a short-range square well potential $U_w(z) =\epsilon $ for $z<\delta$ and
$U_w(z) =0$ otherwise, whereby the range $\delta = 0.125$ (in units of the
maximal bond length extension $l_{max}$ between adjacent beads). The effective
bonded interaction is described by the FENE (finitely extensible nonlinear
elastic) potential.
\begin{equation}
U_{FENE}= -K(1-l_0)^2ln\left[1-\left(\frac{l-l_0}{l_{max}-l_0} \right)^2 \right]
\label{fene}
\end{equation}
with $K=20, l_{max}=1, l_0 =0.7, l_{min} =0.4$

The nonbonded interactions are described by the Morse potential.
\begin{equation}
\frac{U_M(r)}{\epsilon_M} =\exp(-2\alpha(r-r_{min}))-2\exp(-\alpha(r-r_{min}))
\end{equation}
with $\alpha =24,\; r_{min}=0.8,\; \epsilon_M/k_BT=1$.

We use periodic boundary conditions in the $x-y$ directions and impenetrable
walls in the $z$ direction. We study homopolymer chains, regular multiblock
copolymers, and random copolymers (with a fraction of attractive monomers,
$p=0.25,\;0.5,\;0.75$) of length $32$, $64$, $128$, $256$ and $512$. The size of
the box is $64\times 64\times 64$ in all cases except for the $512$ chains where
we use a larger box size of $128\times 128\times 128$. The standard Metropolis
algorithm is employed to govern the moves with  self avoidance automatically
incorporated in the potentials. In each Monte Carlo update, a monomer is chosen
at random and a random displacement attempted with $\Delta x,\;\Delta y,\;\Delta
z$ chosen uniformly from the interval $-0.5\le \Delta x,\Delta y,\Delta z\le
0.5$. The transition probability for the attempted move is calculated from the
change $\Delta U$ of the potential energies before and after the move as
$W=exp(-\Delta U/k_BT)$. As for a standard Metropolis algorithm, the attempted
move is accepted if $W$ exceeds a random number uniformly distributed in the
interval $[0,1)$.

Typically, the polymer chains are originally equilibrated in the MC method
for a period of about $10^6$ MCS (depending on degree of adsorption $\epsilon$
and chain length $N$ this period is varied) whereupon one performs $200$
measurement runs, each of length $8\times 10^6$ MCS. In the case of random
copolymers, for a given composition, i.e., percentage $p$ of the $A-$monomers,
we create a new polymer chain in the beginning of the simulation run by means of
a randomly chosen sequence of segments. This chain is then sampled during the
course of the run, and replaced by a new sequence in the beginning of the next
run.
\begin{figure}[ht]
\begin{center}
\includegraphics[scale=0.32, angle=0]{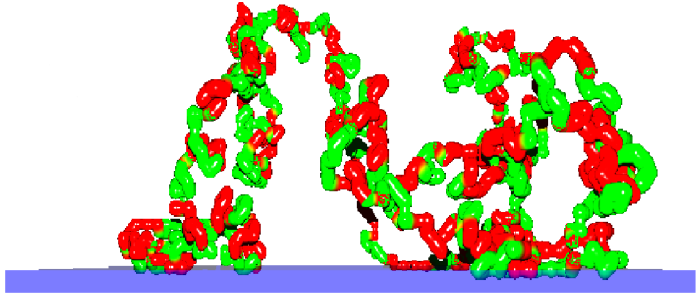}
\caption{Schematic representation of a grafted chain close to criticality.
Snapshot of a regular multiblock copolymer with length $N=2048$ and block
size $M=8$ from the PERM simulation.}
\label{Criticality}
\end{center}
\end{figure}
\subsection{Coarse-grained lattice model with PERM}
\label{ssec:perm}

The adsorption of a diblock $AB$ copolymer with one end (monomer $A$) grafted to
a flat impenetrable surface and with only the $A$ monomers attractive to the
surface is described by self-avoiding walks (SAW) of $N-1$ steps on a simple
cubic lattice with restriction $z \ge 0$. The partition sum may be written as
\begin{equation}
     Z_N^{(1)}(q) =\sum_{N_s} A_N(N_s) q^{N_s}   \label{ZNA}
\end{equation}
where $A_N(N_s)$ is the number of configurations of SAWs with $N$ steps having
$N_s$ sites on the wall, and $q=e^{\epsilon/k_BT}$ is the Boltzmann factor
($\epsilon>0$ is the attractive energy between the monomer $A$ and the wall).
Clearly, any copolymer will collapse onto the wall, if $q$ becomes sufficiently
large. Therefore one expects a phase transition from a grafted but otherwise
detached phase into an adsorbed phase, similar to the transition observed for
homopolymers.

The pruned-enriched Rosenbluth method (PERM)~\cite{Grassberger}, also used in
our simulations, is a {\em biased} chain growth algorithm with resampling
("population control") and depth-first implementation. Polymer chains are built
like random walks by adding one monomer at each step. Thus the total weight of a
configuration for a polymer consisting of $N$ monomers is a product of those
weight gains at each step, i.e. $W_N=\Pi_{i=0}^{N-1} w_i$. As in any such
algorithm, there is a wide range of possible distributions of sampling so we
have the freedom to give a bias at each step while the chain grows, and the bias
is corrected by means of giving a weight to each sample configuration, namely,
$w_i \rightarrow w_i/p_i$ where $p_i$ is the probability for putting the monomer
at step $i$. In order to suppress the fluctuations of weights as the chain is
growing, the population control is done by "pruning" configurations with too low
weight and "enriching" the sample with copies of high-weight configurations.
Therefore, two thresholds are introduced here, $W_n^+=c^+ Z_n$ and $W_n^-=c^-
Z_n$, where $Z_n=\frac{1}{M_n}\sum_{config.} W_n$ from the $M_n$ trail
configuration is the current estimate of partition sum at the $n-1$ step, $c^+$
and $c^-$ are constants of order unity and $c^+/c^- \approx 10$.
   In order to compare with the results obtained by the first MC method, we
simulate homopolymers of length $N=2048$ and multi-block copolymers with
block size $M=2^k\;,$ $k=0,1,2,\cdots,9$ (see Fig. \ref{Criticality}). The
number of monomers is increased
to $N=8192$ as the block size increases. There are $10^5\div 10^6$
independent configurations for each measurement. We also simulate random
copolymers of $N=2048$ monomers with composition $p=0.125$, $0.25$, $0.50$, and
$0.75$.

\section{Scaling Behavior at Criticality}
\label{sec:statics}
\subsection{A homopolymer chain}
\label{ssec:homo}
It is well known~\cite{Eisen,Sommer,Metzger} that a single polymer chain
undergoes a transition from a non-bound into an adsorbed state when the
adsorption energy $\epsilon$ per monomer increases beyond a critical value
$\epsilon_c \approx k_BT$ (where $T$ stands for the temperature of the system).
The adsorption transition can be interpreted as a second-order phase transition
at the critical point (CAP) of adsorption $\epsilon = \epsilon_c$ in the
thermodynamical limit, i.e., $N \rightarrow \infty$. Close to the CAP the number
of surface contacts $N_s$ scales as $N_s (\epsilon = \epsilon_c)\sim N^{\phi}$.
The numerical value of $\phi$ is somewhat controversial and lies in a range
between $\phi = 0.59$ (ref. \cite{Eisen}) and $\phi = 0.484$
(ref.~\cite{Grassberger}), we adopt however the value $\phi = 0.50\pm 0.02$
which has been suggested as the most satisfactory\cite{Metzger} by comparison
with comprehensive simulation results.

How does polymer structure vary with adsorption strength? Consider a chain
tethered to the surface at the one end. The fraction of monomers on the surface
$n = N_s/N$ may be viewed as an order parameter measuring the degree of
adsorption. In the thermodynamic limit $N \rightarrow \infty$, the fraction $n$
goes to zero ($\approx {\cal O}(1/N)$) for $\epsilon \ll \epsilon_c$, then near
$\epsilon_c$, $n \sim N^{\phi - 1}$ whereas for $\epsilon \gg \epsilon_c$ (in
the strong coupling limit) $n$  is independent of $N$. Let us measure the
distance from the CAP by the dimensionless quantity $\kappa = (\epsilon -
\epsilon_c)/\epsilon_c$ and also introduce the scaling variable $\eta \equiv
\kappa N^{\phi}$. The corresponding scaling ansatz\cite{Bhatta_MM08} is then
$n(\eta) = N^{\phi - 1} \: G \left(\eta\right)\;$ with the scaling function
\begin{eqnarray}
G (\eta)=
\begin{cases}
{\rm const} &\quad, \quad {\rm for} \quad \eta \rightarrow 0\\
\eta^{(1 - \phi)/\phi} &\quad, \quad {\rm for } \quad \eta \gg 1
\end{cases}
\label{Scaling_two}
\end{eqnarray}
The resulting scaling behavior of $n$  follows as,
\begin{eqnarray}
n \propto \begin{cases} 1/N &\quad, \quad {\rm for} \quad \kappa \ll 0\\
N^{\phi - 1} &\quad, \quad {\rm for} \quad  \kappa\rightarrow 0\\
\kappa^{(1 - \phi)/\phi} &\quad, \quad {\rm for} \quad \kappa  \gg 1
\end{cases}
\label{Order_param}
\end{eqnarray}
The gyration radius in direction perpendicular to the surface, $R_{g\perp}
(\eta)$, has the form $R_{g\perp} (\eta) = a N^{\nu} {\cal G}_{g\perp} \left(
\eta\right)$. One may determine the form of the scaling function ${\cal
G}_{g\perp}(\eta)$ from the fact that for $\kappa < 0$ one has $R_{g\perp} \sim
a N^{\nu}$ so that ${\cal G}_{g\perp} = {\rm const}$. In the opposite limit,
$\eta \gg 0$ the $N$-dependence drops out and ${\cal G}_{g\perp} (\eta) \sim
\eta ^{-\nu/\phi}$. In result
\begin{eqnarray}
R_{g\perp} (\eta) \propto
\begin{cases}
a N^{\nu} &\quad, \quad {\rm for} \quad \eta \le 0\\
\kappa^{- \nu/\phi} &\quad, \quad {\rm for} \quad \eta \gg 0\;.
\end{cases}
\label{Perp}
\end{eqnarray}
The gyration radius in direction parallel to the surface has similar scaling
representation, $R_{g\parallel} (\eta) = a N^{\nu} {\cal G}_{g\parallel} \left(
\eta\right)$. Again at $\kappa < 0$ the gyration radius $R_{g\parallel} \sim a
N^{\nu}$ and ${\cal G}_{g\parallel} = {\rm const}$. At $\eta \gg 0$ the chain
behaves as a two-dimensional self-avoiding walk (SAW), i.e., $R_{g\parallel}
\sim a N^{\nu_2}$, where $\nu_2=3/4$ denotes the Flory exponent in two
dimensions. In result, the scaling function behaves as ${\cal G}_{g\parallel}
(\eta) = \eta^{(\nu_2 - \nu)/\phi}, \quad {\rm at} \quad \eta \gg 0\;.$ Thus
\begin{eqnarray}
R_{g\parallel} (\eta) \propto \begin{cases} a N^{\nu} &\quad, \quad {\rm at}
\quad \eta \le 0\\
\kappa^{(\nu_2 - \nu)/\phi} N^{\nu_2}&\quad, \quad {\rm at} \quad \eta \gg 0\;.
                   \end{cases}
\label{Parall}
\end{eqnarray}

\begin{figure}[htb]
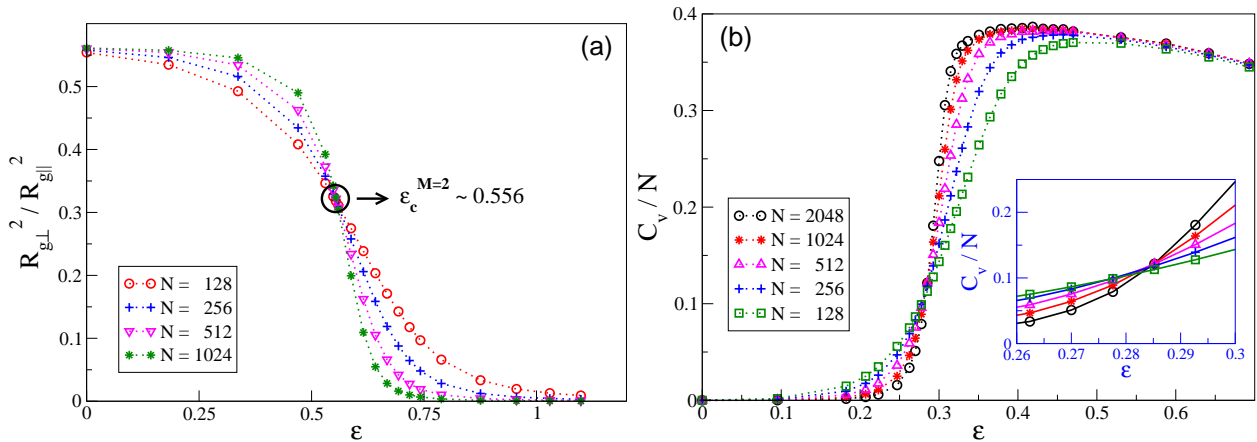

\centering
\includegraphics*[width=.46\textwidth]{fig6b.eps}
\includegraphics*[width=.46\textwidth]{fig5b.eps}
\caption[]{Variation of the ratio $R_{g\perp}^2/R_{g\parallel}^2$ for a
copolymer of block size $M=2$ (a), and of the specific heat per monomer,
$C_v(\epsilon)/N$ for a homopolymer (b) with surface energy $\epsilon$ for
different chain length $N$. One gets $\epsilon_c^{M=2}\approx 0.556$ and
 $\epsilon_c^h \approx 0.284$ (PERM).}
\label{fig:scaling_eps}       
\end{figure}

The study of the ratio $r (\eta) \equiv R_{g\perp}/R_{g\parallel}$ of gyration
radius components is a convenient way to find the value of $\epsilon_c$
\cite{AMKB_Macromolecules1996,Metzger,Sommer}. In fact, from the previous
scaling equations $r(\eta) = {\cal G}_{g\perp}(\eta)/{\cal
G}_{g\parallel}(\eta)$. Hence at the CAP, i.e., at $\eta \rightarrow 0$ the
ratio $r (0) = const.$ is {\em independent} of $N$. Thus, by plotting $r$ vs.
$\epsilon$ for different $N$ all such curves should intersect at a single point
which gives $\epsilon_c$ - cf. Fig.~\ref{fig:scaling_eps}a.

If the well-known picture of blobs~\cite{deGennes} is invoked, then in the limit
$\kappa N^{\phi} \gg 1$ the adsorbed chain can be visualized as a string of {\it
adsorption blobs} which forms a pancake-like quasi-two-dimensional layer on the
surface. The blobs are defined to contain as many monomers $g$ as necessary to
be on the verge of being adsorbed and therefore carry an adsorption energy of
the order of $k_B T$ each. The thickness of the pancake $R_{g\perp}$ corresponds
to the size of the blob while the chain conformation within a blob stays
unperturbed (i.e., it is simply a SAW), therefore $g \sim \left( R_{g\perp} / a
\right)^{1/\nu} = \kappa^{-1/\phi} $ where we have used Eq.~\ref{Perp}. The
gyration radius can be represented thus as
\begin{eqnarray}
R_{g\parallel}  = R_{g\perp} \left( \frac{N}{g}\right)^{\nu_2}
\propto \kappa^{(\nu_2 - \nu)/\phi} N^{\nu_2}
\label{Blob}
\end{eqnarray}
and one goes back to Eq.~\ref{Parall} which proves the consistency of the
adsorption blob picture. Generally speaking, the number of blobs, $N/g \sim
\kappa^{1/\phi} N$, is essential for the main scaling argument in the
above-mentioned scaling functions. The adsorption on a plane at $\kappa > 0$
is due to free energy gain which is then proportional to the number of blobs,
i.e., $F - F_{\rm bulk} \propto - N/g \sim - \kappa^{1/\phi}N$. The
expression for the specific heat per monomer may be then obtained as
\begin{equation}\label{cv}
C_V =- \frac{\partial^2 (F - F_{\rm bulk} )}{\partial^2 \kappa }
\propto \kappa^{-\alpha}\;,
\end{equation}
where $\alpha = 2 - \phi^{-1}$. If $\phi = 0.5$ then $\alpha = 0$ and the
specific heat does not diverge but rather undergoes a jump at the CAP - cf.
Fig.~\ref{fig:scaling_eps}b.

\subsection{Multiblock Copolymer Adsorption}
\label{sec:multiblock}

\begin{figure}[htb]
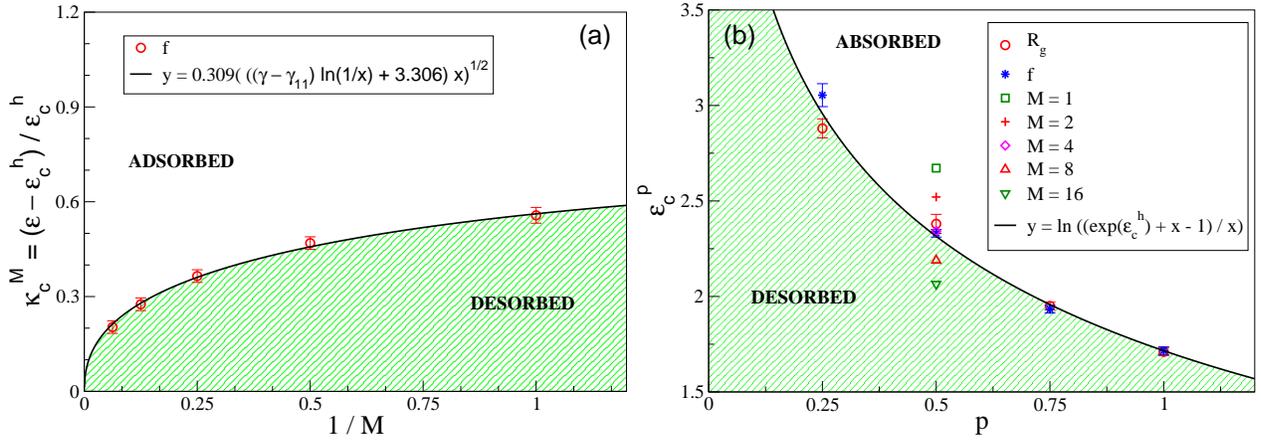

\centering
\includegraphics*[width=.46\textwidth]{fig9a.eps}
\includegraphics*[width=.46\textwidth]{fig12a.eps}
\caption[]{(a) Phase diagram showing $\kappa^M_c =(\epsilon_c^M -\epsilon_c^h) /
\epsilon_c^h$ plotted vs $1/M$ for multiblock copolymers with various values of
$M$ and $\epsilon_c^h=1.716$. The curve gives the best fit of Eq.
\ref{Kappa_vs_M}. (b) The CAP, $\epsilon_c^p$, plotted vs the
composition $p$ for random copolymers. Symbols denote the CAP for multiblock
copolymers with block size $M$. The cureve corresponds to the best fit of Eq.
\ref{Copol}. Data obtained employing  the off-lattice bead-spring model.}
\label{fig:phase_diag}       
\end{figure}

One may now consider the adsorption of a regular multi-block copolymer
comprising
monomers $A$ which attract (stick) to the substrate and monomers $B$
which are neutral to the substrate. In order to treat the adsorption of a
regular multi-block $AB$-copolymer it may be reduced to that of a homopolymer
which has been considered above. Thus a regular multi-block copolymer can be
treated as a ``homopolymer'' where a single $AB$-diblock plays the role of an
{\em effective monomer}~\cite{Corsi}. Let each individual diblock consist of an
attractive $A$-block of length $M$ and a neutral $B$-block of the same length.
Upon adsorption, the $A$-block would form a string of blobs whereas the $B$-part
forms a non-adsorbed loop or a tail. The free energy gain of the attractive
block may be written then (in units of $k_B T$) as $F_{\rm attr} = -
\kappa^{1/\phi} M$ where $\kappa \equiv (\epsilon - \epsilon^h_c)/\epsilon^h_c$
now measures the normalized distance from the CAP $\epsilon^h_c$ of a
homopolymer. The neutral $B$-part which is most frequently a loop connecting
adjacent $A$-blocks, but could also be a tail with the one end free, contributes
only to the entropy loss $F_{\rm rep} = (\gamma - \gamma_{11}) \ln M$ where the
universal exponents $\gamma$ and $\gamma_{11}$ are well known\cite{Vander} (e.g.
in 3$D$ - space $\gamma = 1.159 $, $\gamma_{11} = -0.390$). If a tail is
involved, one should also use the exponent $\gamma_1 = 0.679$ albeit this does
not change qualitatively the expression for $F_{\rm rep}$. These expressions
reflect the standard partition functions for a free chain, a chain with both
ends fixed at a two points, and for a chain, tethered by the one
end~\cite{Vander}. In result the effective adsorption energy of a diblock
is~\cite{Bhatta_MM08}:
\begin{eqnarray}
E (M) = \kappa^{1/\phi} M  - (\gamma - \gamma_{11}) \ln M\;.
\label{Attraction}
\end{eqnarray}
One can tackle the problem of regular copolymer adsorption by mapping it on that
of a ``homopolymer'', consisting of ${\cal N} = N/2M$ such effective units by
using $a \longrightarrow a M^{\nu},\quad \kappa \longrightarrow \Delta =
\frac{E - E^h_c}{E^h_c}, \quad N \longrightarrow {\cal N}$ where $a$ denotes
the monomer size, and $E^h_c$ is the critical adsoption energy of the
renormalized homopolymer. Generally, one would expect $E^h_c$
to be of the order of $\epsilon^h_c$, reflecting the model dependence of the
latter. At the CAP of the multiblock chain one has $\Delta = 0$, thus one can
estimate the deviation $\kappa^M_c$, of the corresponding critical energy of
adsorption, $\epsilon^M_c$, from that of a homopolymer, namely
 \begin{eqnarray}
\kappa^M_c \equiv \frac{\epsilon^M_c - \epsilon^h_c}{\epsilon^h_c}  = \left(
\frac{(\gamma - \gamma_{11}) \ln M + E^h_c}{M}\right)^{\phi}\;.
\label{Kappa_vs_M}
 \end{eqnarray}
where we have used Eq.~\ref{Attraction}. It is seen from the phase diagram,
Fig.~\ref{fig:phase_diag}a, that the deviation $\kappa^M_c$,
Eq.~\ref{Kappa_vs_M} with $\phi = 0.5$, steadily grows with decreasing block
length $M$, in agreement with the computer experiment.
The fraction of effective units on the surface obeys the same scaling law
as given by eq~\ref{Order_param}, i.e., $n \equiv \frac{{\cal N}_s}{{\cal N}} =
{\cal N}^{\phi - 1} G \left( \Delta {\cal N}^{\phi}\right)$ which becomes
accurate, provided (i) $\kappa \ll 1$ but $M \gg 1$ such that $\ln M \gg 1$ and
$\kappa^{1/\phi}M \gg 1$, and (ii) ${\cal N } \gg 1$. Thus, within each
effective unit of the $A$-monomers only $M_s$ will be adsorbed at criticality
whereby $M_s$ scales as $M_s = M^{\phi} G \left( \kappa M^{\phi}\right)$ so
that the total number of adsorbed monomer $N_s = {\cal N}_s  M_s = {\cal N}_s
M^{\phi} G \left( \kappa M^{\phi}\right)$. The adsorbed fraction of monomers
then is expected to scale with both $N$ and $M$ as
\begin{eqnarray}
n \propto N^{\phi - 1} \: G \left( \kappa M^{\phi}\right)
\:  G \left( \Delta \left( \frac{N}{M}\right)^{\phi}\right)\;.
\label{Order_parameter_MB}
\end{eqnarray}
For sufficiently strong adsoption, $\kappa \sqrt{M} \gg 1$ and $\Delta
\sqrt{N/M} \gg 1$, one gets thus $n \propto \kappa \Delta$.

\begin{figure}[htb]
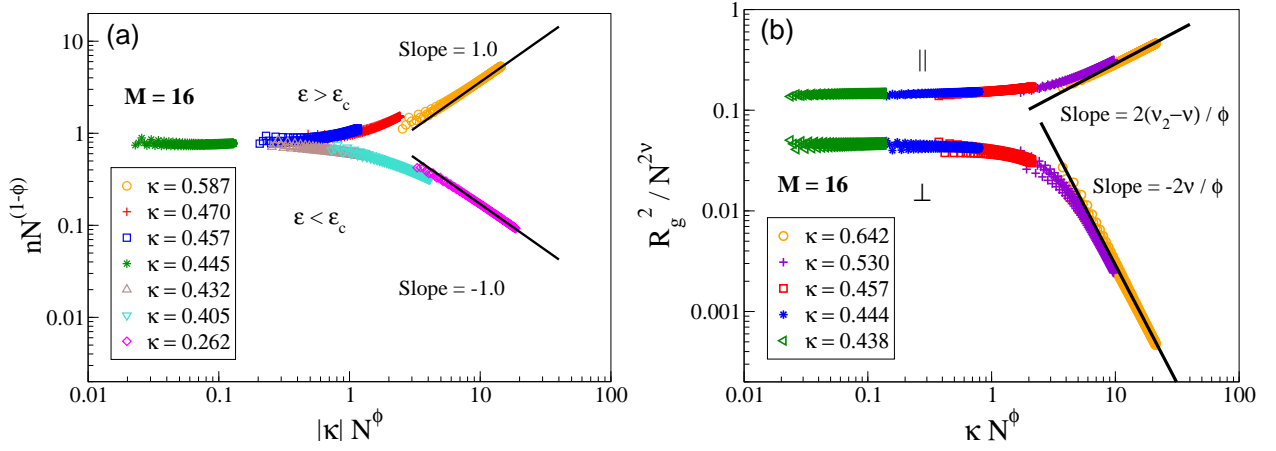

\centering
\includegraphics*[width=.46\textwidth]{fig8c.eps}
\includegraphics*[width=.46\textwidth]{fig8d.eps}
\caption[]{Log-log plots of the order parameter $n$ (a), and
$R_{g\parallel}^2/N^{2\nu}$ and $Rg^2_{g\perp}/N^{2\nu}$ vs $\kappa N^{\phi}$
(b) with $\nu=0.588$ and $\nu_2=3/4$. The straight lines indicate the asymptotic
behaviour of the scaling functions given by Eqs.~(\ref{Order_parameter_MB}),
(\ref{Copoly_perp}), and \ref{Copoly_parallel}   for regular
multi-block copolymers with block size $M=16$ (PERM).}
\label{fig:scaling_MB}       
\end{figure}

The gyration radius
component in direction perpendicular to the surface ${\cal R}_{g\perp} = a
N^{\nu} \: {\cal G}_{g\perp} \left( \Delta \left( \frac{N}{M} \right)^{\phi}
\right)$ becomes ${\cal R}_{\perp} \sim a \Delta^{- \nu/\phi} M^{\nu}$, which
yields
\begin{eqnarray}
{\cal R}_{\perp} \simeq   \frac{a M^{\nu}  {E^h_c}^{2 \nu}}{\left[
\kappa^2 M -(\gamma - \gamma_{11}) \ln M - E^h_c\right]^{2 \nu} }
\label{Copoly_perp}
\end{eqnarray}
In a similar manner, the gyration radius component parallel to the surface has
the form ${\cal R}_{g\parallel} = a N^{\nu} \: {\cal G}_{g\parallel} \left(
\Delta \left( \frac{N}{M}\right)^{\phi}\right)$ which in the limit $\Delta
\sqrt{N/M} \gg 1$ results in ${\cal R}_{g\parallel} \simeq a \left(
\frac{\Delta^{1/\phi}}{M} \right)^{\nu_2 - \nu} \: N^{\nu_2}$, i.e.,
\begin{eqnarray}
{\cal R}_{g\parallel}  \simeq \frac{a \left[ \kappa^2 M - (\gamma
- \gamma_{11}) \ln M - E^h_c\right]^{2(\nu_2 - \nu)} }{M^{\nu_2 - \nu}}
\: N^{\nu_2}\;.
\label{Copoly_parallel}
\end{eqnarray}
As shown in Fig.~\ref{fig:scaling_MB}, one finds indeed the expected scaling
behavior which is demonstrated by the collapse of the simulation data on
few ``master curves'', absorbing cases of different strength of adsorption
$\kappa$.
Like in the homopolymer case, one can define a blob length ${\cal G} \sim
\left({\cal R_{\perp}}/a\right)^{1/\nu} \sim \Delta^{-1/\phi} \: M$ which in the
strong adsorption limit, $\Delta \geq 1$, approaches the block length, ${\cal G}
\simeq M$, as it should be.

\subsection[]{Random Copolymer Adsorption}
\label{sec:random}

The adsorption of a random copolymer on a homogeneous surface has been studied
by Whittington et al.~\cite{Whit_1,Whit_2} within the framework of the annealed
disorder approximation. Physically this means that during the measurements the
chain touches the substrate at random in such a way that one samples all
possible monomers sequences along the backbone of the macromolecule. Following
this assumption~\cite{Whit_1}, let $c_{N}(n)$ be the number of  polymer
configurations such that $n$ units have contact with the surface simultaneously.
The percentage of $A$-monomers (composition) is denoted by $p$. In the annealed
approximation one then averages the partition function over the disorder
distribution, i.e.,
 \begin{eqnarray}
Z (\epsilon) &=& \sum\limits_{n=1}^{N} \sum\limits_{n_p = 0}^{n}
\: c_{N} (n) \: \left( {n \atop n_p}\right)  p^{n_p} (1-p)^{n-n_p}
\: {\rm e}^{\epsilon n_p}\nonumber\\
&=&\sum\limits_{n=1}^{N} c_{N} (n) \: \left[ p {\rm e}^{\epsilon}+1
- p\right]^{n} = \sum\limits_{n=1}^{N} c_{N} (n)
\: {\rm e}^{n \:\epsilon^{h}_{\rm eff}}
\label{Annealed}
\end{eqnarray}
where $\epsilon^{h}_{\rm eff}$ is the attraction energy of an effective
homopolymer. From Eq.~\ref{Annealed} one can see that the annealed problem is
reduced to that of a homopolymer where the effective attractive energy is
defined as $\epsilon^{h}_{\rm eff} = \ln \left[p {\rm e}^{\epsilon} +1 -
p\right]$. Since the homopolymer attraction energy at the CAP
is $\epsilon^{h}_{\rm eff} = \epsilon^{h}_{c}$, the critical attraction energy
$\epsilon=\epsilon_{c}^p$ of a random copolymer will be
\begin{eqnarray}
\epsilon_{c}^p = \ln \left[ \frac{\exp{\epsilon^{h}_{c} + p - 1}}{p}
\right] \geq \epsilon^{h}_{c}
\label{Copol}
\end{eqnarray}
where the composition $0 \leq p \leq 1$. At $p \rightarrow 0$  $\epsilon_{c}^p
\rightarrow \infty$ whereas at $p=1$ $\epsilon_c^p = \epsilon^{h}_{c}$. This
prediction which has been recently confirmed by Monte Carlo
simulations~\cite{Ziebarth}, is plotted in Fig.~\ref{fig:phase_diag}b. It shows
that close to criticality the chain is still rather mobile, so that the sequence
dependence is effectively averaged over the time of the experiment and the of
quenched disorder can be reduced to that of annealed randomness.

\section{Adsorption Kinetics}
\label{sec:kin}

\subsection[]{Variation of the adsorbed fraction with elapsed time - Theory}
\label{ssec:kin_mean}

\begin{figure}[htb]
\centering
\includegraphics*[width=.46\textwidth]{z_n.eps}
\includegraphics*[width=.47\textwidth]{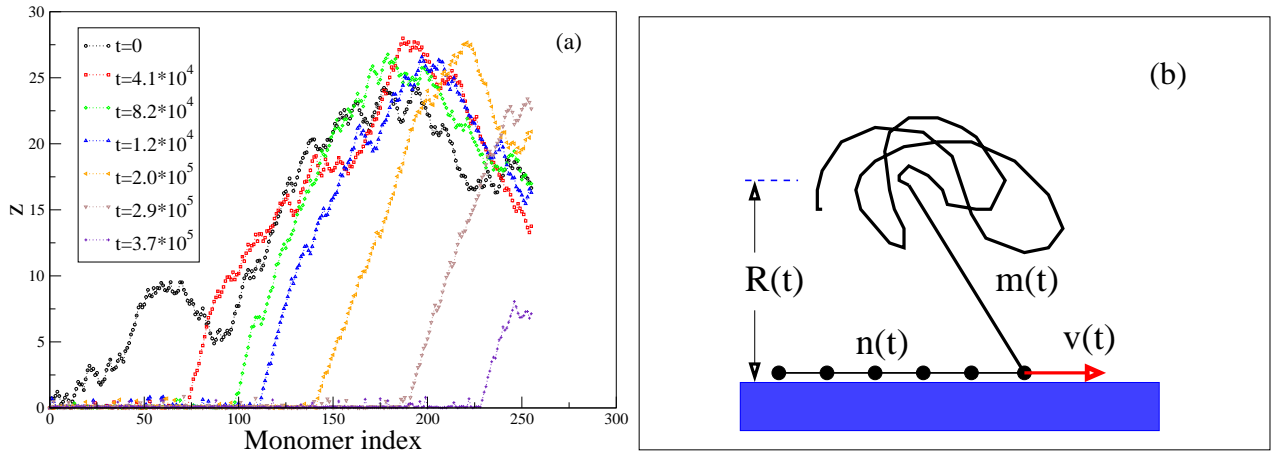}
\caption[]{(a) Chain conformation at successive time moments during the
adsorption process for a polymer with $N=256$. The $z$-coordinate of the $i$-th
monomer  is plotted against monomer index $i$.  (b) Stem-flower picture of the
adsorption dynamics. The total number of adsorbed monomers at time $t$ is
denoted by $n(t)$. The tail which, contains all nonadsorbed monomers, consists
of a stretched part, a ``stem'', of length $m(t)$, and of a nonperturbed part
- a ``flower''. The rate of adsorption is $v(t)$. The distance between the
surface and the front of the tension propagation is $R(t)$.}
\label{fig:stem_flower}       
\end{figure}

We illustrate here how one can use the ``stem - flower'' notion of adsorbing
linear macromolecule, suggested by Descas {\em et al.}~\cite{Descas}, to
describe the observed ``zipping'' dynamics\cite{Ponomar} of adsorption not only
in terms of the average fraction of adsorbed segments but to include also
time-dependent train and tail distribution functions as main constituents of the
dynamic adsorption theory. The simple ``zipping'' mechanism along with the
underlying stem-flower model are illustrated in Fig.~\ref{fig:stem_flower}a,b.
The number of the adsorbed monomers at time $t$ is denoted by $n(t)$. The
nonadsorbed fraction of the chain is subdivided into two parts: a stretched part
("stem") of length $m(t)$, and a remaining part ("flower") which is yet not
affected by the tensile force of the substrate. The tensile force propagation
front is at distance $R(t)$ from the surface. The rate of adsorbtion is denoted
as $v(t)= a \frac{dn(t)}{dt}$, where $a$ is the chain segment length.

A single adsorption event occurs with energy gain $\epsilon$ and entropy loss
$\ln (\mu_3/\mu_2)$, where $\mu_3$ and $\mu_2$ are the connectivity constants in
three and two dimensions, respectively \cite{Vander}, so that the driving free
energy $F_{\rm dr} = \epsilon - k_BT \ln (\mu_3/\mu_2)$ whereas the driving
force $f_{\rm dr} = F_{\rm dr}/a $. The
friction force is related to the pulling of the stem at rate $v(t)$, i.e.,
$f_{\rm fr} = \zeta_0 \: a \: m(t) \: \frac{d n(t)}{d t}$ where $\zeta_0$ is
the Stokes friction coefficient of a single bead. The equation of motion,
following from the balance of driving, and drag force, is then $f_{\rm dr} =
f_{\rm fr}$. Inspection of Fig.~\ref{fig:stem_flower}a suggests that the
distance $R(t)$  between the flower and the plane changes  during the
adsorption process until the flower is eventually ``consumed''. In so doing 
$R (t)$ obeys two relationships:  $R(t) \approx a[n(t)+m(t)]^\nu$ (because it
is actually the size which the chain portion $n (t) + m (t)$ occupied before
the adsorption has started) and  $R(t)\approx m(t)$ (up to a
geometric factor of order unity). Thus $n(t) \approx m(t)^{1/\nu} - m(t)$
which yields $m(t) \approx n(t)^\nu$ for the  typically long stems $m(t) \gg 1$.
From the resulting equation $\zeta_0 \:  n(t)^{\nu} d n(t)/d t = f_{\rm
dr}/a^2$ then follows $n(t) \propto t^{1/(1+\nu)} \approx t^{0.62}$ which is in
good agreement with simulation results\cite{Shaffer,Ponomar,Descas}.

\subsection[]{Time evolution of the distribution functions - Theory}
\label{ssec:kin_PDF}

Consider the instantaneous number of adsorbed monomers $n$ at time $t$, i.e.,
the total train length) distribution function $P(n, t)$. Using the 'Master
Equation' method~\cite{Van_Kampen}, one may derive a system of coupled kinetic
equations for $P(n,t)$ by treating the zipping dynamics as a {\em
one step}  adsorption / desorption process  within an elementary
time interval. Assuming that the corresponding rate constants $w^+(n),\; w^-(n)$
of monomer attachment / detachment are related by the detailed balance
condition~\cite{Van_Kampen} (which is an approximation for a non-equilibrium
process), one can fix their ratio $w^{+}(n-1)/w^{-}(n) = \exp(F_{\rm
dr}/k_BT)$, and even fully specify them by introducing a friction-dependent {\em
transmission} coefficient $q(m) = k_BT/(a^2\zeta) = k_BT/(a^2\zeta_0 m)$
(whereby the stem length $m$ depends on the total train length $n$, according
to $n \approx m^{1/\nu}-m$). Then the one-step Master Equation reads
\cite{Van_Kampen}
\begin{eqnarray}
\frac{d}{d t} \:  P(n, t) &=& w^{-} (n+1)\: P (n+1, t) + w^{+} (n-1) \: P (n-1,
t)\nonumber\\
&-&   w^{+} (n) \: P (n, t) - w^{-}(n) \: P (n, t)\;,
\label{One_Step_ME}
\end{eqnarray}
which along with the boundary conditions
\begin{eqnarray}
\frac{d}{d t} P(1, t) &=& w^{-}(2) P(2, t) - w^{+}(1) P(1, t), \quad
\mbox{for}\: n=1\\ \nonumber
\frac{d}{d t} P(N, t) &=& w^{+}(N-1) P(N-1, t) - w^{-}(N) P(N, t),
\quad \mbox{for}\: n=N
\label{ME_bc}
\end{eqnarray}
and $P(n, t=0) = \delta (n-1)$ fully describe the single chain adsorption
kinetics.

The equation of motion for the mean number of adsorbed segments $\left\langle
n\right\rangle = \sum_{n=1}^{\infty} n P(n, t)$, can be obtained from
Eq.~(\ref{One_Step_ME}), assuming for simplicity $P(N, t) = P(0, t) = 0$:
\begin{eqnarray}
\frac{d}{d t} \: \left\langle n\right\rangle = - \left\langle w^{-} (n)
\right\rangle  + \left\langle w^{+} (n) \right\rangle
\label{First_moment}
\end{eqnarray}
With the relations for the rate constants, $w^+(n),\: w^-(n)$, this
equation of motion becomes
\begin{eqnarray}
 \zeta_0 \: m(t)\: \frac{d}{d t} \:  n(t) = \frac{k_BT}{a^2} \: \left[ 1
- {\rm
e}^{-F_{dr}/k_BT}\right]
\label{Eq_of_motion_3}
\end{eqnarray}
where for brevity we use the notations $n(t) = \left< n \right>$ and $m(t) =
\left< m \right>$. Note that Eq.~(\ref{Eq_of_motion_3}) reduces to the kinetic
equation~\cite{Descas}, derived at the end of Section~\ref{ssec:kin_mean} for
weak driving force, $F_{dr} \ll k_B T$, by neglecting fluctuations in the
zipping
mechanism. Evidently, by taking fluctuations into account, $F_{dr}/a$ is
replaced by a kind of effective {\em second virial} coefficient $(k_B T/a) [1 -
\exp(-F_{dr}/k_BT)]$. Thus, the zipping as a strongly non-equilibrium process
cannot be treated quasistatically by making use a simple ``force balance''.

\subsection[]{Order parameter adsorption kinetics - MC results}
\label{OP_MC}

\begin{figure}[htb]
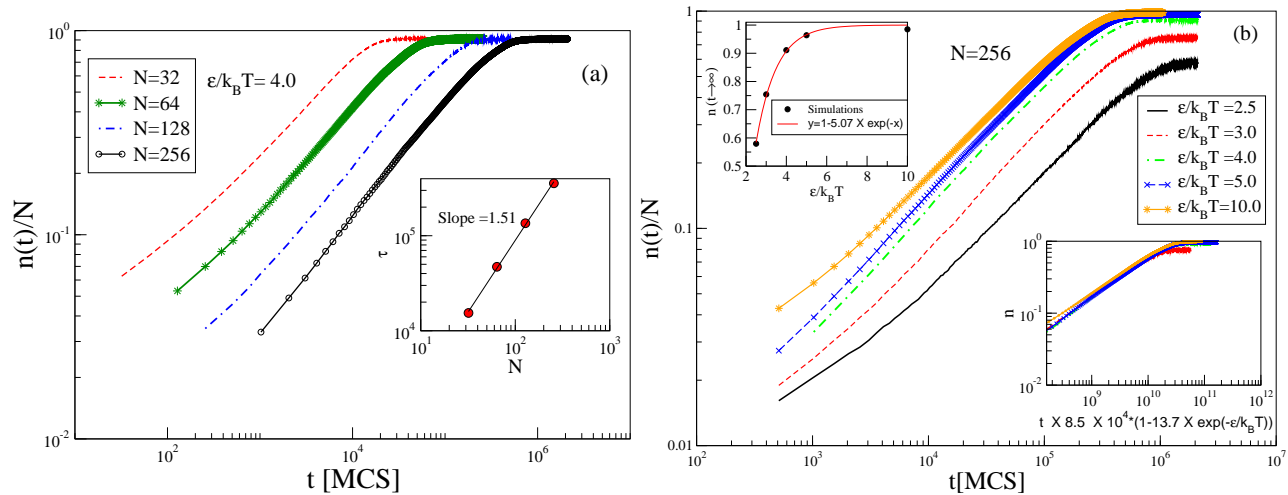

\centering
\includegraphics*[width=.46\textwidth]{OP_t.eps}
\includegraphics*[width=.48\textwidth]{OP_epsilon.eps}
\caption[]{(a) Time evolution of the order parameter (fraction of adsorbed
segments) for four different chain lengths $N=32,\;64,\;128,\; \mbox{and}\; 256$
at surface potential $\epsilon/k_BT=4.0$. The slope of the $N=256$-curve is
$0.56$. The inset shows the scaling of the adsorption time with chain
length,$\tau\propto N^{1.51}$. The time $\tau$ is determined from the
intersection point of the late time plateau with the tangent $t^{0.56}$ to the
respective $n(t)$-curve. (b) Adsorption kinetics for different strengths
$\epsilon$ of the surface potential. The variation of the plateau height (i.e.,
the fraction of adsorbed monomers at equilibrium) with $\epsilon$ is depicted in
the upper inset where the solid line $n_{t\rightarrow \infty}=1-5\exp\left( -
\epsilon/k_BT\right )$ describes the equilibrium number of defects
(vacancies). The lower inset shows a collapse of the adsorption transients on a
single 'master curve', if the time axis is rescaled appropriately.}
\label{fig:MOP}       
\end{figure}

The time variation of the order parameter $n(t)/N$ (the fraction of adsorbed
segments) for homopolymer chains of different length $N$ and strong adhesion
$\epsilon/k_BT=4.0$ is shown in Fig.~\ref{fig:MOP}a,b whereby the observed
straight lines in double-log coordinates suggest that the time evolution of the
adsorption process is governed by a power law. As the chain length $N$
is increased, the slope of the curves grows steadily, and for length $N=256$ it
is equal to $\approx 0.56$. This value is close to the theoretically expected
slope of $(1+\nu)^{-1}\approx 0.62$. The total time $\tau$ it takes a polymer
chain to be fully adsorbed is found to scale with chain length as $\tau\propto
N^{\alpha}$ whereby the observed power ${\alpha \approx 1.51}$ is again somewhat
smaller than the expected one $1+\nu\approx 1.59$, most probably due to
finite-size effects. One may also verify from Fig.~\ref{fig:MOP}b that for a
given length $N$ the final (equilibrium) values of the transients at late times
$t\rightarrow \infty$ grow while the curves are horizontally shifted to shorter
times as the surface potential gets stronger. Nontheless, the slope of
the $n(t)$ curves remains unchanged when $\epsilon/k_BT$ is varied, suggesting
that the kinetics of the process is well described by the assumed zipping
mechanism.
The changing plateau height may readily be understood as reflecting the
correction in the equilibrium fraction of adsorbed monomers due to the presence
of defects (vacancies) for any given value of $\epsilon/k_BT$. For the
transients which collapse on a master curve, cf. the second inset in
Fig.~\ref{fig:MOP}b, one may view the rescaling of the time axis by the
expression $t\rightarrow t [1 - 13.7\exp\frac{-\epsilon}{k_BT}]$ as a direct
confirmation of Eq.~\ref{Eq_of_motion_3} where the time variable $t$ may be
rescaled with the driving force of the process (i.e., with the expression in
square brackets). The factor $\approx 13.7$ gives then the ratio $\mu_3/\mu_2$
of the effective coordination numbers in $3$- and $2$-dimensions of a polymer
chain with excluded volume interactions. $\mu_3$ and $\mu_2$ are model-dependent
and characterize, therefore, our off-lattice model.
\begin{figure}[htb]
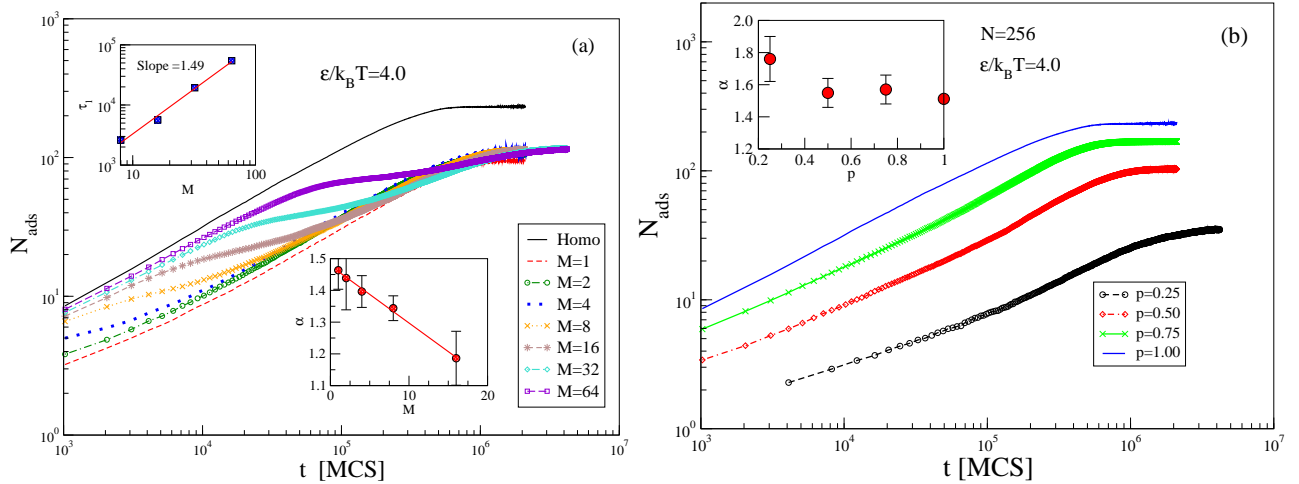

\centering
\includegraphics*[width=.46\textwidth]{MBOP_t.eps}
\includegraphics*[width=.48\textwidth]{OP_random.eps}
\caption[]{(a) Number of adsorbed segments, $N_{ads}(t))$, versus
time $t$ for regular $AB$-copolymers with block size $M=1\div 64$ and length
$N=256$. For comparison, the transient of a homopolymer is shown by
a solid line too. The time interval, taken by the initial
``shoulder'', is shown in the upper left inset. The lower inset
displays the variation of the scaling exponent, $\alpha$, for the
time of adsorption $\tau\propto N^\alpha$ versus block length
relationship. (b) The same as in (a) but for random copolymers of length
$N=256$ and different composition $p=0.25,\; 0.5,\; 0.75$. For $p=1$ one has
the case of a homopolymer. The inset shows the variation of
$\alpha$ with $p$.}
\label{fig:MBOP}       
\end{figure}

The more complex adsorption kinetics, shown in Fig.~\ref{fig:MBOP}a for regular
multiblock copolymers of block size $M$, and in Fig.~\ref{fig:MBOP}b for random
copolymers, suggests however that the power-law character of the order parameter
variation with time is retained except for a characteristic 'shoulder' in the
adsorption transients. Indeed, one should bear in mind that the zipping
mechanism, assumed in our theoretical treatment, is by no means self-evident
when the file of sticking $A$-monomers is interrupted by neutral $B$ segments.
The characteristic shoulder in the adsorption transients of regular multiblock
copolymers manifests itself in the early stage of adsorption and lasts
progressively longer when $M$ grows. The temporal length of this shoulder
reflects the time it takes for a segment from the {\em second} adsorptive
$A$-block in the polymer chain to be captured by the attractive surface, once
the first $A$-block has been entirely adsorbed. For sufficiently large blocks
one would therefore expect that this time interval, $\tau_1$, associated with
the capture event, will scale as the Rouse time, $M^{1+2\nu}$, of a
non-adsorbing tethered chain of length $M$. The observed $\tau_1$ versus $M$
relationship has been shown in the upper left inset in Fig.~\ref{fig:MBOP}a. The
slope of $\approx 1.49$ is less that the Rouse time scaling exponent, $2.18$,
which one may attribute to the rather small values of the block length $M$ that
were accessible in our simulation. One should also allow for scatter in the end
time of the shoulder due to the mismatch in the capture times of all the
successive $A$-blocks in the course of our statistical everaging over many
chains during the computer experiment.

Somewhat surprisingly, $\alpha$ which describes the scaling of the total
adsorption time with polymer size, $\tau \propto N^\alpha$, is observed to {\em
decline} as the block size $M$ is increased - in contrast to the general
trend of regular multiblock copolymers which resemble more and more
homopolymers (where $\alpha = 1+\nu$), as the block size $M \rightarrow
\infty$. Evidently, the frequent disruption of the zipping process for smaller
blocks $M$ slows down the overall adsorption.

In the case of random copolymers, Fig.~\ref{fig:MBOP}b, the transients resemble
largely those of a homopolymer chain with the same number of beads again, apart
from the expected difference in the plateau height which is determined by the
equilibrium number of adsorbed monomers. A rescaling of the vertical axis with
the fraction of sticking monomers, $p$, however, does not lead to coinciding
plateau heights - evidently the loops, whose size also depends on $p$, affect
the equilibrium number of adsorbed monomers. The variation of the observed
scaling exponent $\alpha$ with composition $p$ is shown in the inset to
Fig.~\ref{fig:MBOP}b wherefrom one gets $\alpha \approx 1.6$ with $\alpha$ being
largely independent of $p$. Note that this value is considerably lower than the
power of $2.24$ which has been observed earlier~\cite{Shaffer}, however, for
very short chains with only $10$ sticking beads. One may conclude that even for
random copolymer adsorption the typical time of the process scales as
$\tau\propto N^\alpha$, as observed for homo - and regular block copolymers. It
is conceivable, therefore, that an {\em effective}  zipping mechanism in terms
of renormalized segments, that is, segments consisting of an $A$ and $B$ diblock
unit of length $2M$ for regular multiblock copolymers provides an adequate
notion of the way the adsorption kinetics may be treated even in such more
complicated cases. For random copolymers the role of the block length $M$ would
then be played by the typical correlation length.

\subsection[]{Time evolution of the distribution functions -  MC data}
\label{ssec:kin_PDF_MC}

\begin{figure}[htb]
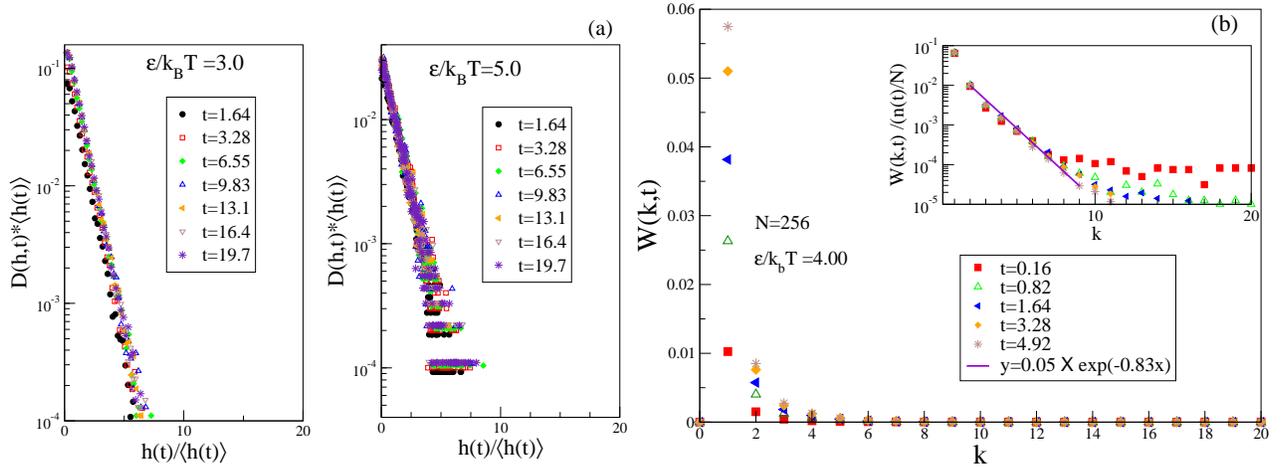

\centering
\includegraphics*[width=.46\textwidth]{PDF_trains.eps}
\includegraphics*[width=.47\textwidth]{PDF_loops.eps}
\caption[]{(a) Distribution of train lengths during the adsorption process of a
homopolymer chain with $N=256$ at two strengths of the adsorption potential
$\epsilon$, shown in semi-log coordinates. PDFs for different times (in units of
$10^5$ MCS) collapse on master curves when rescaled by the mean train length
$h_{av}(t)$. (b) Distribution of loop lengths $W(k,t)$ for $N=256$ and
$\epsilon/k_BT=4.0$ during ongoing polymer adsorption. In the inset the PDF is
normalized by $n(t)$ and shown to be a straight line in log-log coordinates.}
\label{fig:PDF}       
\end{figure}

One gains most comprehensive information regarding the adsorption process from
the time evolution of the different building blocks (trains, loops, and tails)
Probability Distribution Functions (PDF)~\cite{Bhatta_PRE08}. From the MC
simulation data, displayed in Fig.~\ref{fig:PDF}a, for example, one may verify
that the resulting distribution $D(h,t)$ of different train lengths is found to
be exponential, in close agreement with the theoretically expected
shape~\cite{Bhatta_PRE08}, predicted under the assumption that local equilibrium
of loops of unit length is established much faster than the characteristic time
of adsorption itself. When scaled with the mean train length $h_{av}(t) =
\langle h(t)\rangle$, at time $t$, in both cases for $\epsilon/k_BT = 3.0\;
\mbox{and} \;5.0$ one finds an almost perfect straight line in semi-log
coordinates. One may thus conclude that  $D(h,t)$ preserves its exponential form
during the course of the adsorption process, validating thus the conjecture of
rapid local equilibrium. The latter however is somewhat violated for the case of
very strong  adsorption, $\epsilon/k_BT=5.0$, where the  rather  scattered data
suggests that the
process of loop equilibration is slowed down and the aforementioned time
separation is deteriorated.

\begin{figure}[htb]
\centering
\includegraphics*[width=.46\textwidth]{tail_PDF_time.eps}
\includegraphics*[width=.48\textwidth]{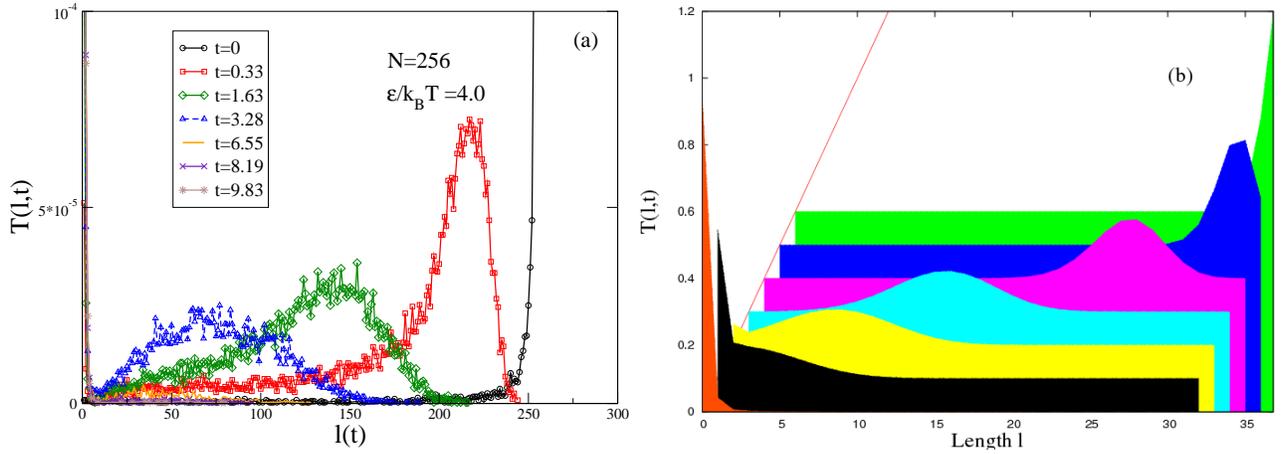}
\caption[]{(a) Distribution of tail size for different times (in
units of $10^5$ MCS) during the polymer chain adsorption for a chain with
$N=256$ at
$\epsilon/k_BT=4.0$. (b) The same as in (a) as derived from the solution of the
ME for chain length $N=32$. For better visibility the time slices for $t=1,\;
5,\; 30\;100,\; 150,\; 200,\; \mbox{and} \;300$ are shifted along the time axis
and arranged such that the initial distribution for $t=1$ is represented by the
most distant slice.}
\label{fig:tail_time}       
\end{figure}

The PDF of loops $W(k,t)$ at different times after the onset of adsorption is
shown in Fig.~\ref{fig:PDF}b. Evidently, the distribution is sharply peaked at
size {\em one} whereas less than the remaining $20\%$ of the loops are of size
two. Thus the loops can be viewed as single thermally activated defects
(vacancies) comprising a desorbed single bead with both of its nearest neighbors
still attached to the adsorption plane. As the inset in Fig.~\ref{fig:PDF}b
indicates, the PDF of loops is also described by an exponential function. The
PDFs for loops at different time collapse on a master curve, if scaled
appropriately with the instantaneous order parameter $n(t)/N$.
Eventually, in Fig.~\ref{fig:tail_time}a we present the observed PDF $T(l,t)$ of
tails for different times $t$ after the start of adsorption, and compare the
simulation results with those from the numeric solution of
Eq.~(\ref{One_Step_ME}), taking into account that $T(l,t) = P(N-l,t)$. One may
readily verify from Fig.~\ref{fig:tail_time} that the similarity between
simulational and theoretic results is really strong. In both cases one starts at
$t=1$ with a sharply peaked PDF at the full tail length $l(t=1) = N$. As time
proceeds, the distribution  becomes broader and its maximum shifts to smaller
values. At late times the moving peak shrinks again and the tail either
vanishes, or reduces to a size of single segment which is expressed by the sharp
peak at the origin of the abscissa.

\section*{Summary}

The main focus of this contribution has been aimed at the adsorption transition
of random and regular multiblock copolymers on a flat structureless substrate
whereby by different means - scaling considerations and computer simulations - a
consistent picture of the macromolecule behavior at criticality is derived.

As a central result one should point out the phase diagram of regular multiblock
adsorption which gives the increase of the critical adsorption potential
$\epsilon^M_c$ with decreasing length $M$ of the adsorbing blocks. For very
large block length, $M^{-1}\rightarrow 0$, one finds that the CAP approaches
systematically that of a homogeneous polymer.

The phase diagram for random copolymers with quenched disorder which gives the
change in the critical adsorption potential, $\epsilon^p_c$, with changing
percentage of the sticking $A$-monomers, $p$, is observed to be in perfect
agreement with the theoretically predicted result which has been derived by
treating the adsorption transition in terms of the ``annealed disorder''
approximation.

Evidently, a consistent picture of how some basic polymer chain properties of
interest such as the gyration radius components perpendicular and parallel to
the substrate, or the fraction of adsorbed monomers at criticality, scale when a
chain undergoes an adsorption transition emerges regardless of the particular
simulation approach. An important conclusion thereby concerns the value of the
universal crossover exponent $\phi=0.5$ which is found to remain unchanged,
regardless of whether homo-, regular multiblock-, or random polymers are
concerned. Thus the universality class of the adsorption transition of a
heteropolymer is the same as that of a homopolymer.

Concerning the adsorption kinetics of a single polymer chain on a flat surface,
it is shown that within the ``stem-flower'' model  and the assumption that the
segment attachment process follows a ``zipping'' mechanism, one may adequately
describe the time evolution of the adsorbed fraction of monomers and of the
probability distribution functions of the various structural building units
(trains, loops, tails) during the adsorption process. For regular multiblock and
random copolymers it is found that the adsorption kinetics strongly resembles
that of homopolymers. The observed deviations from the latter suggest plausible
interpretations in terms of polymer dynamics, however, it is clear that
additional investigations will be warranted before a complete picture of the
adsorption kinetics in this case is established too.
%
%

%
%



\end{document}